\documentclass[aps,twocolumn,groupedaddress]{revtex4}

\usepackage[dvips]{color}
\usepackage{graphicx}
\usepackage{dcolumn}
\usepackage{bm}

\def\be{\begin{equation}}
\def\ee{\end{equation}}

\def\g{g_{l}}

\begin{document}


\title{Failure of the displaced-squeezed state for spin-boson models in the thermodynamic limit}


\author{A. W. Chin}
\email[]{ac307@cam.ac.uk}
\affiliation{Cavendish Laboratory, J.J. Thomson Avenue, Cambridge CB3 0HE, United Kingdom }


\date{\today}
\begin{abstract}
We present an analysis of a variational coherent-squeezed state that has been discussed in the literature as a potential ground state for the spin-boson model. We show that when the system-size scaling of the spin-bath coupling is included properly, all squeezing effects and non-universal physics vanish in the thermodynamic limit. We also present finite-size corrections to the renormalisation of the spin's coherence, showing that squeezing effects are also absent to leading order in the inverse bath-size.
\end{abstract}


\maketitle

\section{Introduction}
The spin-boson model is one of the most important theoretical models for studying dissipation and decoherence in quantum systems, and has been applied to numerous systems in chemistry\cite{parson04}, biology\cite{renger02}, and the emerging fields of quantum computation and quantum devices\cite{makhlin01}. The model is simple, comprising a coherent two level system (TLS) that is coupled to a large bath of harmonic oscillators, yet despite this simplicity, the model is exceptionally rich and continues to provide new insights into open many-body quantum physics.

One particularly important issue, especially for practical quantum devices, is the robustness of the coherence of the TLS in equilibrium with the bath. Numerous studies on this have revealed some extremely interesting physics, including quantum phase transitions between coherent and incoherent TLS states, and unusual temperature dependence of the cohrence\cite{leggett1,weiss,silbey,vojta05,chin06,lu07,bulla03,kehrein96}.

Amongst these studies, the use of trial wave functions has provided some very intuitive insights into the problem. There are two commonly discussed states which are both based on, and improve on, the adiabatic approximation discussed by Leggett et al\cite{leggett1}. The first was that of Silbey and Harris, who modified the adiabatic approximation to allow for simple non-adiabatic responses from the low frequency modes in the problem\cite{silbey}. This state was shown to correctly describe the coherent-incoherent transition for Ohmic coupling\cite{silbey}, and has recently been used to study the transition in the sub-Ohmic system, giving results in excellent agreement with those obtained by other methods\cite{bulla03,vojta05,lu07,chin06}.

The second state is known as the displaced-squeezed state (DSS) and was proposed by Chen, Zhang, and Wong\cite{chen891}. This state is based on two effects; adiabatic displacement of the bath modes, and spatial deformation of the oscillator wave functions. At zero temperature it has been shown that the energy of the DSS can be significantly lower than the Silbey-Harris state for the case of strong coupling between one oscillator and the TLS\cite{chen892}, and this has also been claimed for strong coupling to a super-Ohmic continuum of bath modes\cite{shi94}.

This DSS also predicts coherent-incoherent transitions in Ohmic and sub-Ohmic systems\cite{chen92,dutta94}, but with significantly different critical properties than those found in the rest of the literature. Moreover, these critical properties depend on the explicit form of the TLS-Bath coupling, which is unusual as it has been shown using path integral techniques that the spin-boson physics should be controlled solely by the spectral function of the bath\cite{leggett1,weiss90}. As the DSS is claimed to be the most stable ground state for strong coupling, there has been much discussion of this breakdown of `universality' as a function of coupling strength\cite{chen92,shi94}, a breakdown which we shall show does not exist thermodynamically.

In this paper we use an effective Hamiltonian theory to show that when the system-size scaling of the TLS-bath coupling is included properly, all squeezing and non-universal features of the DSS vanish in the thermodynamic limit. The physics which emerges is simply that of the basic adiabatic approximation. Finite-size effects are then discussed, with no evidence for squeezing found to leading order in the inverse system-size.

\section{The spin-boson model}
\begin{figure}
\includegraphics[width=7cm]{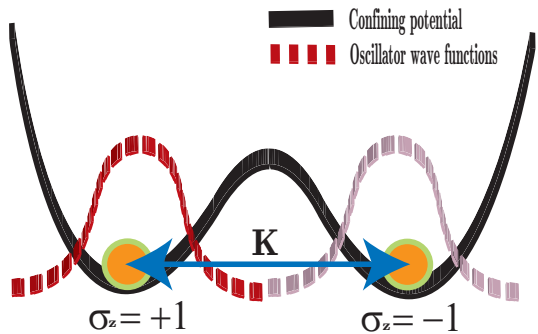}
\caption{The Coherent Two-level system is modeled as particle tunnelling in double well potential. Bath oscillators become correlated with the position of the particle, renormalising the matrix element for tunnelling due to the poor overlap of their displaced wave functions.}
\label{fig1}
\end{figure}
We shall use a standard picture of the spin-boson system in which we consider the TLS as a particle tunnelling between the minima of a double well potential as shown in fig.\ref{fig1}. The Hamiltonian for the spin-boson model is given by\cite{leggett1},

\begin{equation}
H_{sb}=-K\sigma_{x}+\sigma_{z}\sum_{l}g_{l}(a_{l}+a_{l}^{\dagger}) +\sum_{l}\omega_{l}a_{l}^{\dagger}a_{l},\label{hsb}
\end{equation}
where $\sigma_{x,z}$ are Pauli matrices, $a_{l},a^{\dagger}_{l}$ are the bosonic creation and annihilation operators for the bath modes, $g_{l}$ is the coupling between the TLS and the $l$th bath mode, $\omega_{l}$ is the frequency of mode $l$, and $K$ is the tunnelling matrix element between localised states. We have chosen the TLS basis such that the state localised in the left hand well is $|\uparrow\rangle$ and $|\downarrow\rangle$ is the right well state.

Following Leggett et al we assume that the bath is truly macroscopic and that the modes can be treated in the continuum limit with a smooth density of states up to some cut-off frequency $\omega_{c}$\cite{leggett1}. The physics of the spin-boson model is then normally determined solely by the spectral function of the bath $J(\omega)$,

\begin{eqnarray}
J(\omega)&=&\sum_{l}\g^2\delta(\omega-\omega_{l}),\label{delta}\\
&=&\frac{1}{2}\alpha\omega_{s}^{1-s}\omega^{s}\theta(\omega_{c}-\omega).
\label{j}
\end{eqnarray}

The RHS of eq.(\ref{j}) is a phenomenological power-law where $\alpha$ is a dimensionless coupling strength, $\omega_{s}$ is a typical frequency scale of the bath, and $s$ is the exponent of the frequency dependence. The ultra-violet cut-off $\omega_{c}$ for the spectral function will be taken to be the largest energy scale in the problem, much larger than $K$ or $\omega_{s}$. This form for $J(\omega)$ can derived for many specific microscopic interactions, e.g. phonons, E.M modes, and many other physical examples can be found in the literature\cite{weiss}.

The dynamical and thermodynamical behaviour of the TLS depends critically on the coupling strength and the exponent $s$. As a result, three types of bath are distinguished in the literature according to their value of $s$, the super-Ohmic bath ($s>1$), Ohmic bath ($s=1$), and the sub-Ohmic bath ($s<1$)\cite{leggett1}.
\section{Adiabatic approximation and the Displaced-Squeezed state}
In the absence of tunnelling the TLS will be localised in one of the wells, and the spin-boson Hamiltonian can be solved exactly. The two degenerate ground states $|+\rangle,|-\rangle$ are given by,
\begin{eqnarray}
|+\rangle&=&U_{A}|\uparrow\rangle|0\rangle,\hspace{0.5cm}|-\rangle=U_{A}|\downarrow\rangle|0\rangle,\\
U_{A}&=&\exp[-\sigma_{z}\sum_{l}\g\omega_{l}^{-1}(a_{l}-a_{l}^{\dagger})].
\end{eqnarray}
These ground states can be understood intuitively; the localised TLS simply creates a static force which displaces all the bath modes, the displacement being described by the action of the shift operator $U_{A}$ on the vacuum of all bath modes $|0\rangle$. When $K$ is finite the TLS can tunnel between the wells, and no exact solution for the problem is known. However, bath modes with frequencies much higher than the tunnelling rate $2K$ can respond almost instantaneously to the slow tunnelling motion and can be eliminated using an adiabatic approximation\cite{weiss}.

In the zeroth-order adiabatic approximation, where all modes are assumed to follow the TLS instantaneously, the tunnelling leads to a coherent ground state  $|g.s\rangle=\frac{1}{\sqrt{2}}(|+\rangle+|-\rangle)$, characterised by a renormalised coherent level splitting $\tilde{K}=\langle+|K\sigma_{x}|-\rangle$ between the tunnelling states $|+\rangle$ and $|-\rangle$. Physically, the tunnelling probability is reduced by the TLS-bath correlation, as the overlap between the states $|+\rangle$ and $|-\rangle$ is suppressed by the relative displacement of the oscillators as shown in fig.\ref{fig1}. Calculating $\tilde{K}$ explicitly, and using the definition of the spectral function to write the sum over bath modes as an integral, we find that,
\begin{equation}
\tilde{K}=K\exp\left[-\alpha\omega_{S}^{1-s}\int_{0}^{\omega_{c}}\omega^{s-2}\,d\omega\right].
\label{ad}
\end{equation}

From Eq.(\ref{ad}) and Eq.(\ref{j}) we can see that $\tilde{K}$ is finite for super-Ohmic baths, and zero for Ohmic and sub-Ohmic baths due to the infra-red divergence of the integral in Eq. (\ref{ad}). Thus the zero-order adiabatic approximation predicts that the TLS is always localised in an Ohmic or sub-Ohmic environment at $T=0$ K, a result which is known to be incorrect\cite{bulla03,vojta05,chin06,kehrein96}. This failure can be traced to the mistreatment of the slow modes in the problem, for which the adiabatic approximation is clearly not valid.

The DSS of Chen, Zhang and Wu\cite{chen891} attempts to improve on the adiabatic approximation by also allowing for spatial distortion of the oscillator wave functions as they adiabatically follow the TLS. As is discussed in Ref. \cite{chen891}, the distortion of the oscillator wave functions can be described using the generators of bosonic squeezed states, minimum uncertainty states commonly used in quantum optics\cite{loudon}. The effective displaced-squeezed ground state they propose is given by $|\psi_{S}\rangle=U_{A}U_{S}\frac{(|\uparrow\rangle+|\downarrow\rangle)|0\rangle}{\sqrt{2}}$ where $U_{S}$ is given by,

\begin{equation}
U_{S}=\exp[-\sum_{l}\gamma_{l}(a_{l}^{2}-a_{l}^{\dagger\,2})].
\end{equation}
The parameters $\gamma_{l}$ describe the amount of spatial distortion of the oscillator wavefunction, or alternatively, the squeezing of the bosonic quadrature operators\cite{loudon}. Once determined, the squeezing parameters modify the  renormalisation of the tunnelling matrix element by altering the overlap integral of the adiabatically displaced oscillator states. This modification can lead to qualitatively different physical results, including the possibility of finite $\tilde{K}$ for Ohmic and sub-Ohmic systems.

\section{Variational method and the effective Hamiltonian}
We now use an adapted version of the variational method of Silbey and Harris to calculate the effective tunnelling matrix element of the TLS in the displaced-squeezed state at $T=0$ K\cite{silbey}. First we make a canonical transformation to generate the Hamiltonian in the basis of the displaced-squeezed state,  $\tilde{H}=U_{A}U_{S}H_{sb}U_{S}^{-1}U_{A}^{-1}$.
\begin{eqnarray}
&\tilde{H}&=-\hat{K}_{+}\sigma_{+}-\hat{K}_{-}\sigma_{-}+\sum_{l}\omega_{l}\sinh(2\gamma_{l})\nonumber\\
&-&\sum_{l}\g^{2}\omega_{l}^{-1}+\sum_{l}\omega_{l}a_{l}^{\dagger}a_{l}(\cosh^{2}(2\gamma_{l})+\sinh^{2}(2\gamma_{l}))\nonumber\\
&+&\sum_{l}\omega_{l}(a_{l}^{2}+a_{l}^{\dagger 2})\cosh(2\gamma_{l})\sinh(2\gamma_{l}),\label{Ht}\\
\end{eqnarray}
where the operators $K_{+},K_{-}$ obey,
\begin{equation}
\hat{K}_{+}=\hat{K}_{-}^{*}=K\exp\left[-2\g\omega_{l}^{-1}e^{-2\gamma_{l}}(a_{l}-a_{l}^{\dagger})\right].
\label{kil}
\end{equation}

In the approximations discussed previously, the interacting bath-TLS system can be described by an effective non-interacting Hamiltonian characterised by the renormalised tunnelling matrix element $\tilde{K}$. We derive this form of Hamiltonian, as a mean-field approximation to $\tilde{H}$, by introducing the expectation value of the tunnelling operators $\tilde{K}=\langle0|\hat{K}_{+}|0\rangle=\langle0|\hat{K}_{-}|0\rangle$, which has the explicit form,
\begin{equation}
\tilde{K}=K\exp\left[-2\sum_{l}g_{l}^{2}\omega_{l}^{-2}e^{-4\gamma_{l}}\right].
\label{ksum}
\end{equation}
Adding and subtracting $\tilde{K}\sigma_{x}$ to $\tilde{H}$, we then write the Hamiltonian as $\tilde{H}=H_{0}+\tilde{V}$,
\begin{eqnarray}
H_{0}&=&-\tilde{K}\sigma_{x}+\sum_{l}\omega_{l}\sinh(2\gamma_{l})-\sum_{l}\g^{2}\omega_{l}^{-1}\nonumber\\
&+&\sum_{l}\omega_{l}a_{l}^{\dagger}a_{l}[\cosh^{2}(2\gamma_{l})+\sinh^{2}(2\gamma_{l})],\\
\tilde{V}&=&\tilde{H}-H_{0}.
\end{eqnarray}

Following Silbey and
Harris \cite{silbey}, we now compute the Bogoliubov-Feynman upper
bound on the free energy of the system $A_{B}$\cite{feynmanstat}. Bogoliubov's
theorem states that the true free energy $A$ of the model is related to $A_{B}$ by\cite{mazo73},
\begin{equation}
A \leq A_{B},\nonumber
\end{equation}
\begin{equation}
A_{B} = -\beta^{-1}\,\ln \textrm{Tr}\exp(-\beta\,H_{0}) + \langle
\tilde{V} \rangle _{H_{0}}.
\label{ab}
\end{equation}
The angular brackets denote the thermal expectation value calculated with respect to $H_{0}$, and computing the trace using the simple eigenstates of $H_{0}$, we find that $\langle \tilde{V}\rangle_{H_{0}}=0$. The free energy bound at $T=0$ K is thus,
\begin{equation}
A_{B}=-\tilde{K}+\sum_{l}\omega_{l}\sinh(2\gamma_{l})-\sum_{l}\g^{2}\omega_{l}^{-1}.
\end{equation}
We note here that, like all previous studies in the literature, this variational method does not consider the possibility that the separation of the Hamiltonian into $H_{0}+\tilde{V}$ may lead to divergent fluctuations if higher order perturbations are calculated i.e. that the variational state is an inappropriate starting point for analysis of this problem. This matter will be discussed in a forthcoming study of these variational methods.

The variational parameters $\gamma_{l}$ are determined by minimising the energy of the system w.r.t the set $\{\gamma_{l}\}$. We find that $\gamma_{l}$ satisfies,
\begin{equation}
e^{8\gamma_{l}}=1+\frac{8\tilde{K}\g^{2}}{\omega_{l}^{3}},
\label{gam}
\end{equation}
and substituting this into Eq.(\ref{ksum}), we find that $\tilde{K}$ obeys the self-consistent equation,

\begin{equation}
\tilde{K}=K\exp\left[-\sum_{l}\left(\frac{2\g^{2}}{\omega_{l}^{2}}\right)\left(1+\frac{8\tilde{K}\g^{2}}{\omega_{l}^{3}}\right)^{-\frac{1}{2}}\right].
\label{dis}
\end{equation}

This self-consistent equation must be solved to find the effective tunnelling matrix element of the mean field effective Hamiltonian $H_{0}$. For a finite $\tilde{K}$ the ground state is a delocalised, coherent superposition state, whilst for $\tilde{K}=0$ the TLS is incoherent and localised in one of classical well states. Due to the self-consistent nature of Eq.(\ref{dis}), the function $\tilde{K}$ may vanish at a discrete value of the coupling strength $\alpha$, signifying a type of phase transition between coherent and incoherent phases\cite{kehrein96,chin06,lu07,bulla03,chen92}.

The presence of $\g^{2}$ in two places in Eq.(\ref{dis}) means that the renormalisation of $K$ is determined by both the spectral function and $\g^{2}$. In the literature\cite{chen891,chen92,shi94}, $g_{l}$ is taken to have a general power-law form $g_{l}=g_{0}\left(\frac{\omega_{l}}{\omega_{c}}\right)^{n}$, where $g_{0}$ is a constant. However, this form for the coupling constants cannot be valid as it doesn't not take into account the scaling of these constants with system size. From Eq.(\ref{delta}) we see that $J(\omega)$ is the product of the square of the coupling constant at $\omega_{l}=\omega$ and the density of states per unit frequency. As the density of states per unit frequency is proportional to $\mathcal{N}$, where $\mathcal{N}$ is the total number of bath oscillators, the microscopic coupling constants have to scale as $\mathcal{N}^{-\frac{1}{2}}$ to ensure that the spectral function is well defined in the thermodynamic limit. This point is made very clear in the origin review of Leggett et al\cite{leggettnote}.

The non-scaling form for $g_{l}$ used in the DSS literature may be related to the continuous coupling function $g(\omega)$ that can be defined from the spectral function and bath density of states, as discussed by Weiss\cite{weiss}. The spectral function can be written as $J(\omega)=g^{2}(\omega)\mathcal{D}(\omega)$\cite{weiss}, where $\mathcal{D}(\omega)$ is the bath density of states per unit volume. This coupling function $g(\omega)$ can take the non-scaling form used in the DSS literature, as the intrinsic scaling of the microscopic couplings $g_{l}$ has been absorbed into the definition of the density of states per unit volume. However, the quantities that appear in the self-consistent equation Eq. (\ref{dis}) are the actual microscopic coupling constants from the spin boson Hamiltonian, and these must include the system-size scaling. As the dependence of $\tilde{K}$ on $\mathcal{N}$ is essential to our discussion, we assume a general power-law form for the couplings given by,
\begin{equation}
g_{l}^{2}=\frac{\alpha\omega_{c}^{2}}{\mathcal{N}}\left(\frac{\omega_{l}}{\omega_{c}}\right)^{n},
\end{equation}
where $n$ an exponent greater than zero\cite{weiss}.

We now use the definition of the spectral function given in Eq.(\ref{delta}) to write the sum in Eq.(\ref{dis}) as an integral over a continuous distribution of bath modes. In this limit, the self-consistent equation for $\tilde{K}$ takes the form,

\begin{eqnarray}
\tilde{K}&=&K\exp\left[-2\int_{0}^{\omega_{c}}\frac{J(\omega)\omega^{-2}\,d\omega}{\left(1+8\alpha\omega_{c}^{2-n}\tilde{K}\omega^{n-3}\mathcal{N}^{-1}\right)^{\frac{1}{2}}}\right],\nonumber\\
\tilde{K}&=&K\exp\left[-I(\tilde{K})\right]\label{ik}.
\end{eqnarray}
We then write the integral $I(\tilde{K})$ in dimensionless form,
\begin{eqnarray}
I(\tilde{K})&=&\alpha\omega_{s}^{1-s}\int_{0}^{\omega_{c}}\frac{\omega^{s-2}\,d\omega}{\left(1+8\alpha\tilde{K}\omega^{n-3}\omega_{c}^{2-n} \mathcal{N}^{-1}\right)^{\frac{1}{2}}},\nonumber\\
I(\tilde{K})&=&\alpha\left(\frac{\omega_{s}}{\omega_{c}}\right)^{1-s}\int_{0}^{1}\frac{x^{s-2}\,dx}{\left(1+\Delta(\tilde{K}) x^{n-3}\right)^{\frac{1}{2}}},
\label{scaling}
\end{eqnarray}
where we have defined the dimensionless numbers $x=\frac{\omega}{\omega_{c}}$ and $\Delta(\tilde{K})=\frac{8\alpha\tilde{K}}{\mathcal{N}\omega_{c}}$. Comparing Eq.(\ref{scaling}) with Eq.(\ref{ad}) we can see that the inclusion of squeezing effects introduces a square root factor into the integrand which can have a profound effect on the renormalisation of $K$. The squeezing provides an effective infra-red cut-off to the integral, and thus prevents the infra-red divergence if $n<2s+1$ and $\Delta(\tilde{K})$ is finite. Squeezing effects can therefore potentially lead to coherent ground states for Ohmic and sub-Ohmic baths, as well as sharp coherent-incoherent transitions, as is described for the Ohmic case in refs. \cite{chen92,dutta94}.

However as there can only be a true infra-red divergence in the thermodynamic limit, we must consider what happens to the solutions of Eq.(\ref{ik}) as $\mathcal{N}\rightarrow\infty$. In this limit we see that $\Delta\rightarrow 0$, and thus the integral in Eq.(\ref{scaling}) reverts to the form given by the adiabatic approximation\cite{leggett1,weiss}. Therefore for Ohmic and sub-Ohmic baths, the DSS always leads to an incoherent thermodynamic ground state, which from Eq.(\ref{gam}) leads self-consistently to the vanishing of the squeezing parameters $\gamma_{l}$.  Interestingly, a recent variational study of squeezing effects in a weakly interacting Bose-Einstein condensate also found that the squeezing vanishes in the thermodynamic limit.\cite{haque06}.

We can also show that this conclusion is independent of when we choose to take the thermodynamic limit by solving the self-consistent equation at finite $\mathcal{N}$ and then sending $\mathcal{N}\rightarrow\infty$. At finite $\mathcal{N}$ there can be no phase transition as there will always be an infra-red cut-off $\omega_{IR}$ arising from the finite-size of the system. This frequency natural scales inversely with the linear dimensions of the system, and can be written $\omega_{IR}=\mathcal{N}^{-\frac{1}{3}}\omega_{0}$, where $\omega_{o}$ is some system dependent frequency which depends on the density of oscillators $\mathcal{N}/V$.  The squeezing effects provide an effective infra-red cut-off at a frequency of approximately $\omega_{c}\Delta^{\frac{1}{3-n}}\propto\mathcal{N}^{-\frac{1}{3-n}}$, which vanishes faster than $\omega_{IR}$ as $\mathcal{N}$ becomes very large. Thus in the limit of large $\mathcal{N}$, the squeezing factor can be ignored in the integrand of Eq. (\ref{scaling}), and the renomalisation is given by the adiabatic formula of Eq.(\ref{ad}), but with an infra-red cut-off at $\omega_{IR}$. The renormalised matrix element is then given by,
\begin{equation}
\tilde{K}=K\exp\left[-\frac{\alpha}{1-s}\left(\frac{\omega_{s}\mathcal{N}^{\frac{1}{3}}}{\omega_{0}}\right)^{1-s}\right],
\end{equation}
for the case of sub-Ohmic coupling, and
\begin{equation}
\tilde{K}=K\left(\frac{\omega_{0}}{\omega_{c}\mathcal{N}^{\frac{1}{3}}}\right)^{\alpha},
\end{equation}
for Ohmic coupling.

As expected, for finite $\mathcal{N}$ the TLS retains it's coherence for arbitrary coupling strength, with the coherence vanishing in the thermodynamic limit. Due to the fact that the infra-red divergence is prevented by $\omega_{IR}$ rather than $\omega_{c}\Delta^{\frac{1}{3-n}}$, the coherence does not depend on the TLS-bath coupling, although these finite-size corrections are non-universal as they depend on the bath frequency $\omega_{0}$.

Although we find no evidence for squeezing even at finite $\mathcal{N}$, squeezing effects can be very important for coupling to a single mode\cite{chen892}, or possibly a few discrete modes. For instance, such squeezing effects have recently been predicted for a TLS coupled to a single nanomechanical oscillator\cite{xue07}.

In conclusion, when one uses the correct system-size scaling of the TLS-bath couplings, analysis of the DSS shows that all the squeezing and non-universal effects vanish in the thermodynamical limit. In thermodynamic equilibrium, this variational state thus reproduces the results of the basic adiabatic approximation, and fails to capture the interesting transition physics of Ohmic and sub-Ohmic systems. Squeezing effects are also found to be absent when one considers the leading order finite-size corrections around the zero-order adiabatic state.

\bibliography{decoherence}
\end{document}